\documentclass[12pt]{article}
\usepackage{a4wide,cite,float,latexsym}

\newcommand{\cF}{{\cal F}}
\newcommand{\cO}{{\cal O}}

\newcommand{\nn}{\nonumber}

\newcommand{\eqn}[1]{(\ref{#1})}

\newcommand{\gev}{\mbox{\rm GeV}}
\newcommand{\DeKP}{\Delta_{K\pi}}
\newcommand{\SiKP}{\Sigma_{K\pi}}
\newcommand{\msmall}[1]{\mbox{$#1$}}

\newcommand{\smvs}{\vbox{\vskip 8mm}}

\newcommand{\fgev}{\footnotesize{GeV}}

\newcommand{\newsection}[1]{\section{#1}\setcounter{equation}{0}}

\begin{document}

\begin{titlepage}

\phantom{x} \vspace{-2cm}
\begin{flushright}
{\small\sf IFIC/03-57} \\[-1mm]
{\small\sf FTUV/04-0112} \\[-1mm]
{\small\sf HD-THEP-03-59} \\[-1mm]
{\small\sf FERMILAB-PUB-03-415-T} \\[16mm]
\end{flushright}

\begin{center}
{\LARGE\bf Order \boldmath{$p^6$} chiral couplings from \\[2mm]
 the scalar \boldmath{$K\pi$} form factor}\\[12mm]

{\normalsize\bf Matthias Jamin${}^{1}$, Jos\'e Antonio Oller${}^{2}$
 and Antonio Pich${}^{3}$} \\[4mm]

{\small\sl ${}^{1}$ Institut f\"ur Theoretische Physik, Universit\"at
           Heidelberg,} \\
{\small\sl Philosophenweg 16, D-69120 Heidelberg, Germany}\\
{\small\sl ${}^{2}$ Departamento de F\'{\i}sica, Universidad de Murcia,
           E-30071 Murcia, Spain} \\
{\small\sl ${}^{3}$ Departament de F\'{\i}sica Te\`orica, IFIC,
           Universitat de Val\`encia -- CSIC,}\\
{\small\sl Apt. Correus 22085, E-46071 Val\`encia, Spain} \\[12mm]
\end{center}

{\bf Abstract:}
Employing results from a recent determination of the scalar $K\pi$ form factor
$F_0^{K\pi}$ within a coupled channel dispersion relation analysis \cite{jop01},
in this work we calculate the slope and curvature of $F_0^{K\pi}(t)$ at zero
momentum transfer. Knowledge of the slope and curvature of the scalar $K\pi$
form factor, together with a recently calculated expression for $F_0^{K\pi}(t)$
in chiral perturbation theory at order $p^6$, enable to estimate the $\cO(p^6)$
chiral constants $C_{12}^r(M_\rho)=(0.3 \pm 5.4)\cdot 10^{-7}$ and
$(C_{12}^r+C_{34}^r)(M_\rho)=(3.2 \pm 1.5)\cdot 10^{-6}$. Our findings also
allow to estimate the contribution coming from the $C_i$ to the vector form
factor $F_+^{K\pi}(0)$ which is a crucial ingredient for a precise
determination of $|V_{us}|$ from $K_{l3}$ decays. Our result
$F_+^{K\pi}(0)|_{C_i^r}=-\,0.018 \pm 0.009$, though inflicted with large
uncertainties, is in perfect agreement with a previous estimate by Leutwyler
and Roos already made twenty years ago.

\vfill



\end{titlepage}

\newsection{Introduction}

Chiral Perturbation Theory ($\chi$PT) \cite{we79,gl84,gl85a,ec95,pi95}
provides a very powerful framework to study the low-energy dynamics of the
lightest pseudoscalar octet. After having been developed to order $p^4$ in
the energy expansion in the fundamental papers by Gasser and Leutwyler
\cite{gl84,gl85a}, the increasing demand for higher precision in the
low-energy description of QCD suggested to extend this expansion to the next
order $p^6$ \cite{bcegs97,bce99,bce00}.

However, the predictive power of $\chi$PT decreases when one tries to increase
the accuracy, because the chiral symmetry constraints are less powerful at
higher orders. While the number of allowed operators is 10 at order $p^4$,
parameterised by the chiral constants $L_i^r$, it already grows to 90 at the
next order $p^6$. Nevertheless, the situation is not as hopeless as it might
seem, since to a given physical observable, only a few of the chiral couplings
contribute, thus in certain cases allowing to determine all appearing couplings
from phenomenology.

One such set of observables are the strangeness-changing form factors which
parametrise the weak $K\pi$ transition amplitude. The vector form factor
$F_+^{K\pi}(t)$ plays a crucial role in the description of $K_{l3}$ decays,
whereas the scalar form factor $F_0^{K\pi}(t)$ corresponds to the S-wave
projection of the $K\pi$ transition matrix element. At order $p^4$, both
form factors were already calculated almost twenty years ago by Gasser and
Leutwyler \cite{gl85b}.

The value $F_+^{K\pi}(0)$ is an indispensable ingredient in the determination
of the quark-mixing matrix element $|V_{us}|$ from $K_{l3}$ decays, and
therefore good knowledge of this quantity is required in order to determine
$|V_{us}|$ with high precision. Very recently, the calculation of $F_+^{K\pi}$
and $F_0^{K\pi}$ has thus been extended to the next order $p^6$
\cite{ps02,bt03}, and it was demonstrated that at this order only two new
chiral couplings, $C_{12}^r$ and $C_{34}^r$, contribute to $F_+^{K\pi}(0)$.
In addition, it was shown that precisely the same couplings also appear in
the slope and curvature of the scalar form factor $F_0^{K\pi}(t)$. As was
first pointed out in ref.~\cite{bt03}, this would allow for a determination
of the two needed couplings if $F_0^{K\pi}(t)$ was known well enough.

Actually, also recently in a different context, the scalar form factor
 $F_0^{K\pi}(t)$ has been determined for the first time from a dispersive
coupled-channel analysis of the $K\pi$ system \cite{jop01}. As an input in
the dispersion integrals, S-wave $K\pi$ scattering amplitudes were used which
had been extracted from fits to the $K\pi$ scattering data in the framework
of unitarised $\chi$PT with explicit inclusion of resonance fields \cite{jop00}.
The initial motivation to calculate $F_0^{K\pi}(t)$ was the fact that it
determines the strangeness-changing scalar spectral function, which was then
employed to calculate the mass of the strange quark from a QCD sum rule
analysis \cite{jop02}.

Thus we are now in a position to employ the results of ref.~\cite{jop01} for
an estimation of the chiral couplings $C_{12}^r$ and $C_{34}^r$. In section~2,
we briefly review the required expressions for the vector and scalar $K\pi$
form factors and in section~3, based on our previous work \cite{jop01},
we then calculate the slope and the curvature of the scalar form factor
$F_0(t)$. Furthermore, our results for the slope and curvature of $F_0(t)$
are employed to present an estimate of the contributions to the vector form
factor at zero momentum transfer $F_+(0)$, resulting from the order $p^6$
chiral constants, and in section~4, we end with some concluding remarks.
We have also included an appendix in which we present an analytical approach
to the numerical analysis followed in section~3 and discuss additional
alternatives to determine $F_+(0)$.

\newsection{\boldmath{$K\pi$} form factors}

In the Standard Model, the decay of K mesons into a pion and a lepton pair
($K_{l3}$ decay) is mediated by the strangeness changing vector current
$\bar s\gamma_\mu u$. The corresponding hadronic matrix element, which
parametrises the decay $K^0\to\pi^-l^+\nu_l$ has the general form
\begin{equation}
\label{Kl3ME}
\langle\pi^-(p')|\bar s\gamma_\mu u|K^0(p)\rangle \;=\;
(p+p')_\mu\,F_+^{K\pi}(t) + (p-p')_\mu\,F_-^{K\pi}(t) \,,
\end{equation}
where $t=(p-p')^2$. In the following, we shall work in the isospin limit,
and thus the matrix element in eq.~\eqn{Kl3ME} is equal to the corresponding
one which describes the decay $K^+\to\pi^0l^+\nu_l$, up to a global
normalisation factor.\footnote{Isospin breaking corrections resulting from
both order $e^2$ as well as $(m_u-m_d)$ terms have been calculated in
\cite{cknrt01}, and need to be included for a complete phenomenological
analysis of $K_{l3}$ decays.} Therefore, different charge states for kaon
and pion will not be distinguished, and to further simplify the notation,
below we shall also drop the superscript on the form factors and set
$F_\pm(t) \equiv F_\pm^{K\pi}(t)$.

The form factor $F_+(t)$ is also referred to as the vector form factor,
because it specifies the P-wave projection of the crossed-channel matrix
element $\langle 0|\bar s\gamma_\mu u|K\pi\rangle$. The corresponding
S-wave projection is described by the scalar form factor
\begin{equation}
\label{F0}
F_0(t) \;\equiv\; F_+(t) + \frac{t}{(M_K^2-M_\pi^2)}\,F_-(t) \,.
\end{equation}
At order $p^4$ in $\chi$PT, both the vector as well as the scalar form factor
were calculated by Gasser and Leutwyler in \cite{gl85b}. The corresponding
expressions can be found in the original paper, and will not be repeated here.

At the next order $p^6$, both form factors were calculated very recently in
refs.~\cite{ps02,bt03}.\footnote{The diagonal $\pi\pi$ and $K K$ form factors
have been also computed to order $p^6$ in
refs.~\cite{gm91,bct98,bt02,fkm02,bd03}.}
However, the two calculations used different forms of the order $p^6$ chiral
Lagrangian, and therefore it is difficult to compare the results. A comparison
was attempted in ref.~\cite{bt03}, and differences in some parts of the results
were found, but at present no definite conclusions are reached. Awaiting a
clarification of these issues, we decided to employ the more recent analysis
\cite{bt03}, which is based on the formulation of the order $p^6$ chiral
Lagrangian presented in \cite{bce99}.

The value of the vector form factor at $t=0$, $F_+(0)$, plays a crucial role
in the determination of the Cabibbo-Kobayashi-Maskawa (CKM) or quark mixing
matrix element $V_{us}$ from $K_{l3}$ decays \cite{lr84}. Thus, for a high
precision determination of $V_{us}$ from $K_{l3}$ it is mandatory to know the
value of $F_+(0)$ as accurately as possible, since already at the moment the
experimental and theoretical uncertainties to $V_{us}$ are of the same
magnitude. With the upcoming new information on $K_{l3}$ from KLOE
\cite{kloe03} and NA48 \cite{na4803}, actually the uncertainty on $F_+(0)$
will become the limiting factor in the $V_{us}$ determination.

The $\chi$PT result at order $p^6$ for $F_+(0)$ presented in \cite{bt03}
was found to take the following form:
\begin{equation}
\label{Fp0}
F_+(0) \;=\; 1 + \Delta(0) - \frac{8}{F_\pi^4}\,(C_{12}^r+C_{34}^r)\DeKP^2\,,
\end{equation}
where $\DeKP\equiv M_K^2-M_\pi^2$, and $\Delta(0)$ is the correction
which arises from order $p^4$ and $p^6$, but is independent of the order 
$p^6$ chiral constants $C_i^r$. The order $p^4$ chiral constants $L_i^r$ only
appear at $\cO(p^6)$, and a numerical value, based on recent fit results for
the $L_i^r$, was given in \cite{bt03}:
\begin{equation}
\label{Del0}
\Delta(0) \;=\; -\,0.0080 \pm 0.0057\,[{\rm loops}] \pm 0.0028\,[L_i^r] \,.
\end{equation}
It should be pointed out that both, the chiral couplings $C_{12}^r$ and
$C_{34}^r$ as well as $\Delta(0)$, depend on the chiral renormalisation scale.
The value \eqn{Del0} of \cite{bt03} corresponds to the scale $\mu=M_\rho$,
and we shall adopt this choice for the rest of our work. Eq.~\eqn{Fp0}
demonstrates, that the value of $F_+(0)$ only depends on the particular
combination of the two $\cO(p^6)$ chiral constants $(C_{12}^r+C_{34}^r)$.

The expression for the scalar form factor $F_0(t)$ at order $p^6$ in $\chi$PT,
on the other hand, reads:
\begin{equation}
\label{F0t}
F_0(t) \;=\; F_+(0) + \overline\Delta(t) + \frac{(F_K/F_\pi-1)}{\DeKP}\,t +
\frac{8}{F_\pi^4}\,(2C_{12}^r + C_{34}^r)\,\SiKP t -
\frac{8}{F_\pi^4}\,C_{12}^r\,t^2 \,.
\end{equation}
Here, $\SiKP\equiv M_K^2+M_\pi^2$, and $\overline\Delta(t)$ is a function
which receives contributions from order $p^4$ and $p^6$, but like $\Delta(0)$
it is independent of the $C_i^r$, and the order $p^4$ chiral constants $L_i^r$
only appear at order $p^6$. Again, a fit to $K_{e3}^0$ and $K_{e3}^+$ decay
data was presented in ref.~\cite{bt03}. A good fit over the entire phase
space $0\leq t\leq 0.13$ ($t$ in $\gev^2$) is given by
\begin{equation}
\label{delbar}
\overline\Delta(t) \;=\; \overline\Delta_1\,t + \overline\Delta_2\,t^2 +
\overline\Delta_3\,t^3 + {\cal O}(t^4) \;=\;
- \,0.259(9)\,t + 0.840(31)\,t^2 + 1.291(170)\,t^3 \,.
\end{equation}
Our errors on the expansion coefficients $\overline\Delta_i$ have been
estimated from the fit differences to the two $K_{e3}$ channels, and from
the uncertainty due to different sets of $L_i^r$, which at $t=0.13\,\gev^2$
was found to be around $0.0013$ \cite{bt03}.

As should be obvious from eq.~\eqn{F0t}, the relation $F_0(0)=F_+(0)$, which
immediately follows from the definition \eqn{F0}, is satisfied. In addition,
up to order $p^6$, also the full scalar form factor $F_0(t)$ only depends on
the two $\cO(p^6)$ chiral couplings $C_{12}^r$ and $C_{34}^r$, with different
dependencies on the couplings at linear and quadratic order in $t$. Thus, if
the $t$-dependence of the scalar form factor would be known from experiment
or theory, the two couplings $C_{12}^r$ and $C_{34}^r$ could be determined,
enabling us to also predict a value for $F_+(0)$. In the next section, we
will show that such an analysis is actually possible employing a recent
determination of the scalar $K\pi$ form factor $F_0(t)$ from a dispersive
coupled-channel analysis of the $K\pi$ system \cite{jop01}.

\newsection{Scalar \boldmath{$K\pi$} form factor and \boldmath{$F_+(0)$}}

The scalar $K\pi$ form factor has been obtained recently in ref.~\cite{jop01}
from a coupled-channel dispersion-relation analysis. The S-wave $K\pi$
scattering amplitudes which are required in the dispersion relations were
available from a description of S-wave $K\pi$ scattering data in the framework
of unitarised $\chi$PT with resonances \cite{jop00}. The dominant uncertainties
in the scalar $K\pi$ form factor are due to two integration constants which
emerge while solving the coupled channel dispersion relations.

The two integration constants can be fixed by demanding values for $F_0(0)$
as well as $F_0(\DeKP)$. Since $F_0(0)=F_+(0)$, for this input we can invoke
the most recent result of \cite{bt03}, $F_+(0)=0.976\pm 0.010$. Of course,
the value of $F_+(0)$ at order $p^6$ in $\chi$PT also depends on the chiral
couplings $C_{12}^r$ and $C_{34}^r$ which we aim to determine. Therefore, in
the end our determination can be viewed as a consistency check that the
resulting value for $F_+(0)$ is compatible with the input used for $F_0(0)$
in the calculation of the scalar $K\pi$ form factor \cite{jop01}. In order to
fix the second integration constant, we also require a value for $F_0(\DeKP)$,
which, to a very good approximation, is equal to $F_K/F_\pi$ \cite{dw69}:
\begin{equation}
\label{delCT}
F_0(\DeKP) \;=\; \frac{F_K}{F_\pi} + \Delta_{{\rm CT}} \,.
\end{equation}
The correction $\Delta_{{\rm CT}}$ is of order $m_u$ or $m_d$ and has been
estimated to be $\Delta_{{\rm CT}}=-\,3\cdot 10^{-3}$ within $\chi$PT at the
next-to-leading order \cite{gl85b}.

\begin{table}[htb]
\begin{center}
\begin{tabular}{cccc}
\hline
$F_0(0)$ & $F_0(\DeKP)$ & $F_0'(0)$ [\fgev${}^{-2}$] &
$F_0''(0)$ [\fgev${}^{-4}$] \\
\hline
      & 1.21 & 0.804 & 1.674 \\
0.966 & 1.22 & 0.837 & 1.745 \\
      & 1.23 & 0.871 & 1.815 \\
\hline
      & 1.21 & 0.770 & 1.603 \\
0.976 & 1.22 & 0.804 & 1.674 \\
      & 1.23 & 0.837 & 1.744 \\
\hline
      & 1.21 & 0.737 & 1.532 \\
0.986 & 1.22 & 0.770 & 1.603 \\
      & 1.23 & 0.804 & 1.673 \\
\hline
\end{tabular}
\end{center}
\caption{Average values $F_0'(0)$ and $F_0''(0)$ for the unitarised chiral
plus K-matrix fits (6.10K2--4) and (6.11K2--4) of ref.~\cite{jop01}, for
three values of $F_0(0)$ as well as $F_0(\DeKP)$.
\label{tab1}}
\end{table}

The description of the scalar $K\pi$ form factor of ref.~\cite{jop01} now
allows to calculate the first and second derivatives of the scalar form factor
at zero momentum transfer. The different fits to the S-wave $K\pi$ scattering
data have already been discussed in detail in \cite{jop01,jop00} and average
results for the derivatives are presented in table \ref{tab1} for three
different values of $F_0(0)$ as well as $F_0(\DeKP)$. The variation with
respect to the different fits is very minor and has therefore not been displayed
explicitly. The dominant uncertainties stem from the used ranges for $F_0(0)$
and $F_0(\DeKP)$.  We also observe that the variation of $F_0(0)$ and
$F_0(\DeKP)$ in the ranges given above leads to the same uncertainty for both
derivatives $F_0'(0)$ and $F_0''(0)$. Adding these two uncertainties in
quadrature, we obtain:
\begin{equation}
\label{dF1}
F_0'(0) \;=\; 0.804 \pm 0.048 \,\gev^{-2} \,,
\quad\qquad
F_0''(0) \;=\;  1.67 \pm 0.10 \,\gev^{-4} \,.
\end{equation}

The same physical content as the derivative $F_0'(0)$ can also be represented
in two other constants, the scalar $K\pi$ squared radius as well as the slope
parameter $\lambda_0$. From our value for $F_0'(0)$ of eq.~\eqn{dF1}, we
then find
\begin{equation}
\langle r_{K\pi}^2\rangle \;=\; 6\,\frac{F_0'(0)}{F_0(0)}
\;=\; (0.192\pm 0.012)\;{\rm fm}^2 \,, \quad
\lambda_0 \;=\; M_\pi^2\,\frac{F_0'(0)}{F_0(0)} \;=\; 0.0157 \pm 0.0010 \,.
\end{equation}
These results are in perfect agreement to the results by Gasser and Leutwyler
obtained in $\chi$PT at $\cO(p^4)$, $\langle r_{K\pi}^2\rangle=(0.20\pm 0.05)
\,{\rm fm}^2$ and $\lambda_0=0.017\pm 0.004$, though about a factor of four
more precise. On the other hand, the recent finding by Yndur{\'a}in,
$\langle r_{K\pi}^2\rangle=0.31\pm 0.06$ \cite{ynd03}, being $2\sigma$ higher,
is not supported by our result. In ref.~\cite{ynd03}, it was argued that the
larger value found there arises due to the presence of the light $\kappa$
resonance. However, also in our fits to the S-wave $K\pi$ scattering data
\cite{jop00}, a dynamically generated light resonance, which can be identified
with the $\kappa$, was found. Thus the approach of \cite{ynd03} to parametrise
the $\kappa$ resonance with an effective Breit-Wigner form appears
controversial \cite{oll03b}.

On the experimental side, the situation about the slope parameter $\lambda_0$
is rather confusing. For the two decay modes $K^0_{\mu3}$ as well as
$K^\pm_{\mu3}$, the most recent Particle Data Group averages are given by
\cite{PDG02}:
\begin{equation}
\label{la0exp}
\lambda_0 \;=\; \biggl\{ \begin{array}{ccc}
\; 0.030\pm 0.005 & \quad (S=2.0) \quad & \qquad [K^0_{\mu3}] \\
\; 0.004\pm 0.009 & \quad (S=1.8) \quad & \qquad [K^\pm_{\mu3}]
\end{array} \,,
\end{equation}
being in clear disagreement with each other.\footnote{After completion of
our work, we became aware of a very recent high statistics study of the
$K^-\to\pi^0\mu^-\nu$ decay \cite{Yushchenko:2003}, which is in good agreement
to our result of eq.~(3.3).} Furthermore, up to now, in most extractions of
$\lambda_0$ only a linear function was fitted to the data. As shown in
\cite{bt03} for the vector form factor, the inclusion of a curvature term
could produce a sizeable shift in the slope parameter. The average of the
two values \eqn{la0exp} would be compatible with our findings, but in view of
the inconsistence, at present we shall disregard the experimental information
on $\lambda_0$.

The results for $F_0'(0)$ and $F_0''(0)$ of eq.~\eqn{dF1} can now be employed
in order to determine the couplings $C_{12}^r$ as well as $(C_{12}^r+C_{34}^r)$
appearing in the order $p^6$ chiral Lagrangian. We prefer to calculate the
combination $(C_{12}^r+C_{34}^r)$, rather than $C_{34}^r$ itself, because
precisely this combination appears in the $\cO(p^6)$ contribution to $F_+(0)$.
Comparing eq.~\eqn{F0t} with the Taylor expansion for $F_0(t)$ around $t=0$,
one finds the following two relations:
\begin{eqnarray}
\label{c12}
C_{12}^r &=& \Big[\, 2\overline\Delta_2 - F_0''(0) \,\Big]
\frac{F_\pi^4}{16} \,, \\
\label{c12c34} \smvs
(C_{12}^r+C_{34}^r) &=& \Biggl[\, F_0'(0) - \overline\Delta_1 -
\frac{(F_K/F_\pi-1)}{\Delta_{K\pi}} - \frac{8\Sigma_{K\pi}}{F_\pi^4}\,C_{12}^r
\,\Biggr]\frac{F_\pi^4}{8\Sigma_{K\pi}} \,.
\end{eqnarray}
Inserting the given values for $F_0''(0)$ and $\overline\Delta_2$ into
eq.~\eqn{c12}, we obtain the following estimate for $C_{12}^r$:
\begin{equation}
\label{c12r}
C_{12}^r(M_\rho) \;=\; (0.3 \pm 3.3 \pm 4.3)\cdot 10^{-7} \;=\;
(0.3 \pm 5.4)\cdot 10^{-7} \,,
\end{equation}
where the first error corresponds to the variation of $F_0(0)$ and the second
to the remaining uncertainties. Separate results for the three different inputs
for $F_0(0)$ are also given in table~\ref{tab2} below. The huge uncertainty on
$C_{12}^r$ results from the fact that there is an almost complete cancellation
between the two terms in eq.~\eqn{c12}.

Our result of eq.~\eqn{c12r} for $C_{12}^r$ can be directly compared with
an estimate given in ref.~\cite{bt03}, based on assuming that the scalar
pion form factor is dominated by the lowest lying scalar resonance:
\begin{equation}
\label{c12rSMD}
C_{12}^r|_{\rm SMD} \;=\; -\,\frac{F_\pi^4}{8M_S^4} \;\approx\;
-\,1.0\cdot 10^{-5} \,.
\end{equation}
For this estimate it was assumed that the lowest lying scalar resonance can
be identified with the $a_0(980)$. However, recently there appears mounting
evidence that the $a_0(980)$ is of dynamical origin and that the lowest
preexisting scalar resonance is in fact the $a_0(1450)$
\cite{eef86,oo97,oo98,eef02,cenp03,oll03a,bet03}. Furthermore, the estimate of
eq.~\eqn{c12rSMD} does not take into account the scale dependence of the chiral
coupling $C_{12}^r$. In general, the scale dependence of the $C_i^r(\mu)$ can
be deduced from ref.~\cite{bce00}, and is found to be
\begin{equation}
\label{runCi}
C_i^r(\mu_2) \;=\; C_i^r(\mu_1) - \frac{1}{(4\pi)^4} \biggl( \Gamma_i^{(2)}
\ln^2\frac{\mu_1}{\mu_2} + (4\pi)^2 \Big(2\,\Gamma_i^{(1)} +
\Gamma_i^{(L)}(\mu_1)\Big) \ln\frac{\mu_1}{\mu_2} \biggr) \,,
\end{equation}
with the coefficients $\Gamma_i^{(2)}$, $\Gamma_i^{(1)}$ and
$\Gamma_i^{(L)}(\mu)$ being presented in table~II of \cite{bce00}.

Although the scale of the lowest-resonance approximation is not determined, it
appears natural to assume that the relevant scale is close to the scalar mass
$M_S$. Employing $M_S=1.45\,\gev$ and evolving $C_{12}^r$ to $M_\rho$, we
obtain $C_{12}^r|_{\rm SMD}(M_\rho)= 4.0\cdot 10^{-6}$, somewhat smaller and
with opposite sign compared to eq.~\eqn{c12rSMD}. For comparison, the
corresponding result for $M_S=1\,\gev$ would be
$C_{12}^r|_{\rm SMD}(M_\rho)= -\,7.8\cdot 10^{-6}$, close to the estimate
\eqn{c12rSMD}. Complete consistency of the resonance estimate and our result
of eq.~\eqn{c12r} would be obtained for a scalar mass around $M_S=1.25\,\gev$.
From this observation, we conclude that the scalar meson dominance
approximation for the $\cO(p^6)$ chiral constant $C_{12}^r$ is compatible with
our findings, but in view of the strong scale dependence, we are unable to
draw more definite conclusions.

Now, we have all the quantities needed for the determination of
$(C_{12}^r+C_{34}^r)$ from eq.~\eqn{c12c34}. For the ratio $F_K/F_\pi$, we
have employed the value $F_K/F_\pi=1.22\pm 0.01$ from ref.~\cite{lr84}.
Furthermore, our result \eqn{dF1} for the derivative $F_0'(0)$ is required as
an input. As discussed above, half of the uncertainty on this value is given by
the variation of $F_0(\DeKP)$. On the other hand, because of eq.~\eqn{delCT},
the values for $F_K/F_\pi$ and $F_0(\DeKP)$ are strongly correlated and this
correlation should be taken into account for our determination of
$(C_{12}^r+C_{34}^r)$. What we have then done to estimate the uncertainty was
to take half the error on $F_0'(0)$ as given in \eqn{dF1} to be 100\%
correlated with $F_K/F_\pi$, but added the remaining half due to $F_0(0)$
fully uncorrelated. With this treatment of uncertainties, we arrive at the
main result of our work:
\begin{equation}
\label{c12rc34r}
(C_{12}^r+C_{34}^r)(M_\rho) \;=\; (3.2 \pm 1.4 \pm 0.6 )\cdot 10^{-6} \;=\;
(3.2 \pm 1.5)\cdot 10^{-6} \,.
\end{equation}
Again, the first error corresponds to the variation of $F_0(0)$ and the second
to the remaining parameters. Separate values for the three inputs for $F_0(0)$
are also listed in table~\ref{tab2} below.

Also our result of eq.~\eqn{c12rc34r} can be compared directly with the
scalar-resonance estimate of $(C_{12}^r+C_{34}^r)$. Employing the corresponding
expression for $C_{34}^r$ \cite{cnp04},
\begin{equation}
\label{c34rSMD}
C_{34}^r|_{\rm SMD} \;=\; \frac{17}{64}\,\frac{F_\pi^4}{M_S^4} \,,
\end{equation}
and evolving the result to the scale $M_\rho$, we find
$(C_{12}^r+C_{34}^r)|_{\rm SMD}(M_\rho)=5.8\cdot 10^{-6}$. Thus, in this case,
the resonance estimate is $1.7\sigma$ larger than our result of
eq.~\eqn{c12rc34r}, but in view of the strong scale dependence, which is also
present for the combination $(C_{12}^r+C_{34}^r)$, the difference should be
considered as an error estimate of the scalar-resonance approximation. However,
as will be discussed further below, demanding consistency between our input
value for $F_0(0)$ and the resulting output for $F_+(0)$, inspection of table
\ref{tab2} provides some indication that the true value for
$(C_{12}^r+C_{34}^r)(M_\rho)$ might be somewhat larger than our result
\eqn{c12rc34r}, although the large uncertainties make it impossible to draw
more definite conclusions.

As a cross check, our results of eqs.~\eqn{c12r} and \eqn{c12rc34r} can be
used to verify if the value for $\Delta_{{\rm CT}}$ in $\chi$PT at $\cO(p^6)$
is compatible with the order $p^4$ result given above. Evaluating $F_0(t)$ of
eq.~\eqn{F0t} at $t=\DeKP$, one finds
\begin{equation}
F_0(\DeKP)-\frac{F_K}{F_\pi} \;=\; \Delta(0) + \overline\Delta(\DeKP) +
16\,\frac{M_\pi^2}{F_\pi^4}\,\DeKP (2C_{12}^r+C_{34}^r) \;=\;
-\,0.006 \pm 0.007 \,,
\end{equation}
which is in reasonable agreement to the value for $\Delta_{{\rm CT}}$ give
above. However, one should emphasise that it is not clear whether the expansion
of eq.~\eqn{delbar} can still be trusted for $\overline\Delta(\DeKP)$.

Our result of eq.~\eqn{c12rc34r} for $(C_{12}^r+C_{34}^r)$ can readily be
translated into an estimate of the order $p^6$ contribution to $F_+(0)$
resulting from the chiral constants $C_i^r$:
\begin{eqnarray}
\label{Fp0Ci}
F_+(0)|_{C_i^r} &=& -\,\frac{8}{F_\pi^4}\,(C_{12}^r+C_{34}^r)\DeKP^2 \\
\smvs
& & \hspace{-2cm} \;=\; \Biggl[\, \frac{(F_K/F_\pi-1)}{\Delta_{K\pi}} +
\overline\Delta_1 - F_0'(0) + \Big(\overline\Delta_2-\msmall{\frac{1}{2}}
F_0''(0)\Big)\SiKP \,\Biggr]\frac{\DeKP^2}{\SiKP} \;=\;
-\,0.018 \pm 0.009 \nn \,.
\end{eqnarray}
This result is in perfect agreement with an estimate of the same contribution
already given in the pioneering work by Leutwyler and Roos \cite{lr84},
$F_+(0)|_{C_i^r}=-\,0.016 \pm 0.008$. Inspection of table~\ref{tab2} shows
that in this case the uncertainty is dominated by the variation of $F_0(0)$.
The remaining parameters only give a small contribution to the error.

\vspace{4mm}
\begin{table}[htb]
\begin{center}
\begin{tabular}{crccc}
\hline
$F_0(0)$ & $C_{12}^r\;[10^{-7}]$ & $C_{12}^r+C_{34}^r\;[10^{-6}]$ &
$F_+(0)|_{C_i^r}$ & $F_+(0)$ \\
\hline
0.966 & $-\,3.0\pm4.3$ & $4.6\pm0.6$ & $-\,0.026\pm0.003$ & $0.966\pm0.007$ \\
0.976 &    $0.3\pm4.3$ & $3.2\pm0.6$ & $-\,0.018\pm0.003$ & $0.974\pm0.007$ \\
0.986 &    $3.5\pm4.3$ & $1.7\pm0.6$ & $-\,0.009\pm0.003$ & $0.983\pm0.007$ \\
\hline
\end{tabular}
\end{center}
\caption{Results for the different quantities calculated in this work for
three different inputs for $F_0(0)$. The errors correspond to a variation
of all other input parameters.
\label{tab2}}
\end{table}


\newsection{Conclusions}

Employing results of a recent determination of the scalar $K\pi$ form factor
$F_0(t)$ within a coupled channel dispersion relation approach \cite{jop01},
in this work we were able to calculate the slope and the curvature of $F_0(t)$
at zero momentum transfer. Our corresponding results have been given in
eq.~\eqn{dF1}.

Rather recently, the vector and scalar $K\pi$ form factors have also been
calculated in chiral perturbation theory at order $p^6$ in the chiral
expansion \cite{ps02,bt03}. Comparing the resulting expressions for the slope
and curvature of $F_0(t)$, together with our findings, we were in a position
to estimate the order $p^6$ chiral constants $C_{12}^r$ and
$(C_{12}^r+C_{34}^r)$ with the result:
\begin{equation}
C_{12}^r(M_\rho) \;=\; (0.3 \pm 5.4)\cdot 10^{-7} \,,
\quad\qquad
(C_{12}^r+C_{34}^r)(M_\rho) \;=\; (3.2 \pm 1.5)\cdot 10^{-6} \,,
\end{equation}
where the large uncertainties in $C_{12}^r$ are due to numerical cancellations
between the two terms in the relation \eqn{c12}.

The vector form factor at zero momentum transfer $F_+(0)$ \eqn{Fp0} is a
crucial ingredient in the determination of the CKM matrix element $|V_{us}|$
from $K_{l3}$ decays and the $\cO(p^6)$ contribution resulting from the chiral
constants $C_i^r$ happens to be just proportional to the combination
$(C_{12}^r+C_{34}^r)$. Employing our estimate for $(C_{12}^r+C_{34}^r)$,
we then obtained
\begin{equation}
\label{Fp0C}
F_+(0)|_{C_i^r} \;=\; -\,0.018 \pm 0.009 \,,
\end{equation}
being in perfect agreement with an estimate of the same contribution already
given in the original work by Leutwyler and Roos \cite{lr84},
$F_+(0)|_{C_i^r}=-\,0.016 \pm 0.008$. Further improvement of the presented
analysis would require the measurement of $F_0'(0)$, or equivalently
$\lambda_0$, to better than 5\%. This would then also allow to improve the
value of $F_0(\DeKP)$ from our dispersion relation approach to $F_0(t)$, and
thereby to acquire independent information on the value of $F_K/F_\pi$.
Vice versa, also an improvement of our knowledge on the ratio $F_K/F_\pi$
from other sources would help to reduce the uncertainty on $F_+(0)$.

Compiling the information presented in reference \cite{bt03} and this
work, we are in a position to present an updated estimate for $F_+(0)$:
\begin{eqnarray}
\label{Fp0t}
F_+(0) &=& 1 - 0.0227\,[p^4] + 0.0113\,[p^6\mbox{-loops}] + 0.0033\,
[p^6\mbox{-}L_i^r] - 0.018\,[p^6\mbox{-}C_i^r] \nn \\
&& \hspace{3.1mm} \pm\,0.0057\,[\mbox{loops}] \pm 0.0028\,[L_i^r] \pm
0.009\,[C_i^r] \nn \\
&=& 0.974 \pm 0.011 \,,
\end{eqnarray}
where all errors have been added in quadrature. Let us note that while using
the same input parameters as in \cite{ps02}, the authors of ref.~\cite{bt03}
found numerical agreement for the order $p^6$ loop plus $L_i^r$ contribution,
implying that this piece is reasonably well established.

In table~\ref{tab2}, we have again presented our results for $F_+(0)$, for
the three values of $F_0(0)$ separately. We observe, that for the value
$F_0(0)=0.966$, complete agreement between input and output is obtained,
which seems to indicate that this value of $F_0(0)$ is preferred. This would
correspond to a slightly larger value for $(C_{12}^r+C_{34}^r)(M_\rho)$, in
better agreement with the scalar resonance saturation estimate presented in the
last section. However, in view of the large uncertainties, we are unable to
draw further conclusions from this observation. Furthermore, in the analysis
presented above, isospin violation has been neglected for simplicity.
Nevertheless, for a complete phenomenological analysis of $K_{l3}$ decays, it
is mandatory to include isospin violating corrections, as they are crucial to
explain the differences between  $K_{l3}^0$ and $K_{l3}^+$ decays
\cite{lr84,cknrt01,cnp04}. We intend to return to these questions in the
future.

Nevertheless, already at this level a qualitative discussion of the influence
of our results can be given. In the original work by Leutwyler and Roos
\cite{lr84}, the order $p^6$ contribution corresponding to our result
\eqn{Fp0C}, was considered to be the total correction at this order. However,
as was also pointed out in ref.~\cite{bt03}, adding the two-loop correction
as well as the $\cO(p^6)$ contribution proportional to the $L_i^r$, a partial
cancellation takes place and the full $\cO(p^6)$ correction turns out to be
smaller. This in turn implies, that our final result \eqn{Fp0t} for $F_+(0)$
is larger than the corresponding value originally employed in \cite{lr84},
$F_+(0)=0.961$ (already including a tiny isospin correction), and the resulting
value for $|V_{us}|$ from $K_{l3}$ decays should be smaller than the present
Particle Data Group average \cite{PDG02}.\footnote{Depending on the treatment
of the experimental $K_{e3}$ data, also larger values for $|V_{us}|$ can be
obtained \cite{cnp04}. Furthermore, larger values can be accommodated in the
framework of generalised $\chi$PT \cite{fks00}.}

It will be very interesting to see how the upcoming improvements in the
determination of $|V_{us}|$ from $K_{l3}$ decays, both on the theoretical as
well as the experimental side will compare to the determination of $|V_{us}|$
from hadronic $\tau$ decays into strange particles \cite{gjpps03,jam03},
which with upcoming more precise experimental results on the relevant $\tau$
decay rate should also be extremely promising. This should also shed light on
the question of a possible violation of unitarity in the first row of the CKM
matrix.

\smallskip
\subsection*{Acknowledgements}
M.J. is indebted to the Fermilab Theory Group for support and warm hospitality
expressed during a visit where most of this work has been performed. Fermilab
is operated by Universities Research Association Inc. under Contract
No.~DE-AC02-76CH03000 with the U.S. Department of Energy.
This work has also been supported in part by the European Union RTN Network
EURIDICE Grant No. HPRN-CT2002-00311 (J.A.O. and A.P.) as well as by MCYT
(Spain) Grants No. FPA2002-03265 (J.A.O.) and FPA-2001-3031 (A.P. and M.J.).
M.J. would like to thank the Deutsche Forschungsgemeinschaft for support.

\newpage
\appendix{}
\newsection{Analytic dependence on $F_0(0)$ and $F_0(\DeKP)$}

In this appendix, we derive analytical formulae which explicitly show the
dependence of $C_{12}^r$, $C_{12}^r+C_{34}^r$ and $F_+(0)$ on the values of
$F_0(0)$ and $F_0(\DeKP)$. Furthermore, we discuss the dependence of our
results on the input taken for $F_K/F_\pi$.

In ref.~\cite{jop01}, the $K\pi$ and $K\eta'$ scalar form factors were obtained
by numerically solving the so called Muskhelishvili-Omn\`es problem
\cite{musk,basde}. According to these references the most general scalar form
factor can be expressed in terms of two linearly independent solutions
$\{\cF_{10}(s),\cF_{11}(s)\}$ and $\{\cF_{20}(s),\cF_{21}(s)\}$, where the
first subscript indicates the independent solution and the second the channel,
$0$ for $K\pi$ and $1$ for $K\eta'$, so that:
\begin{equation}
\label{newfs}
F(s) \;=\; \alpha_1\, \cF_1(s) + \alpha_2\, \cF_2(s) \,,
\end{equation}
where only the first subscript is indicated, and $F(s)$ is a column vector of
the two form factors $F_0(s)$ and $F_1(s)$ for $K\pi$ and $K\eta'$ channels,
respectively. Generally, $\alpha_{1,2}$ are polynomials \cite{musk} although
in our case they are just constants since the canonical solutions $\cF_i(s)$
vanish at infinity like $1/s$, and we require the resulting scalar form factor
$F(s)$ to also vanish at infinity. The solutions $\cF_i(s)$ are just an output
from the employed T-matrices of ref.~\cite{jop00}, once two normalisation
conditions for each $\cF_i(s)$ are imposed. We choose:
\begin{equation}
\label{newcc}
\cF_{10}(0) \;=\;1 \,,\quad \cF_{20}(0) \;=\; 0 \,,\quad
\cF_{10}(\Delta_{K\pi}) \;=\; 0 \,, \quad \cF_{20}(\Delta_{K\pi}) \;=\; 1 \,.
\end{equation}
With this choice, $F(s)$ in eq.~\eqn{newfs} can be expressed as follows:
\begin{equation}
\label{newfscc}
F(s) \;=\; F_1(0) \, \cF_1(s)+F_1(\Delta_{K\pi})\,\cF_2(s) \,.
\end{equation}
Taking into account the previous expression and eqs.~\eqn{c12} and
\eqn{c12c34}, we then find:
\begin{eqnarray}
\label{ac12}
C_{12}^r &=& \frac{F_\pi^4}{16}\Big[\, 2\overline\Delta_2 -
F_0(0)\cF_{10}''(0)-F_0(\DeKP)\cF_{20}''(0) \,\Big] \,, \\
\label{ac12c34} \smvs
C_{12}^r+C_{34}^r &=& \frac{F_\pi^4}{8\Sigma_{K\pi}}\Biggl[\,
F_0(0)\cF_{10}'(0)+F_0(\DeKP)\cF_{20}'(0)- \overline\Delta_1 -
\frac{(F_K/F_\pi-1)}{\Delta_{K\pi}} - \frac{8\Sigma_{K\pi}}{F_\pi^4}\,C_{12}^r
\,\Biggr] \,, \nonumber \\
\end{eqnarray}
where we have made use of eq.~\eqn{newfscc} to express $F_0'(0)$ and $F_0''(0)$
in terms of the constants $F_0(0)$ and $F_0(\DeKP)$. We can substitute the last
expression for $C_{12}^r$ into eq.~(\ref{ac12c34}), so that:
\begin{eqnarray}
\label{aa.6}
C_{12}^r+C_{34}^r&=&\Biggl[\, F_0(0)\cF_{10}'(0)+F_0(\DeKP)\cF_{20}'(0)
- \overline\Delta_1 - \Sigma_{K\pi}\overline\Delta_2-\frac{(F_K/F_\pi-1)}
{\Delta_{K\pi}} \nonumber \\
&+&\frac{\Sigma_{K\pi}}{2} \Big(F_0(0)\cF_{10}''(0)+F_0(\DeKP)
\cF_{20}''(0)\Big)\,\Biggr]\frac{F_\pi^4}{8\Sigma_{K\pi}} \,.
\end{eqnarray}
This equation, together with eq.~\eqn{ac12}, explicitly shows the dependence
of the chiral counterterms $C_{12}^r$ and $C_{12}^r+C_{34}^r$ on the inputs
$F_0(0)$ and $F_0(\DeKP)$. On the other hand, making use of eq.~\eqn{Fp0},
we can also write:
\begin{eqnarray}
\label{neweqf0}
F_+(0)&=&1+\Delta(0)-\Biggl[\, F_0(0)\left\{\cF_{10}'(0)+
\frac{\Sigma_{K\pi}}{2} \cF_{10}''(0)\right\}+F_0(\DeKP)\left\{\cF_{20}'(0)+
\frac{\Sigma_{K\pi}}{2}\cF_{20}''(0)\right\} \nonumber \\
&-& \overline\Delta_1 - \Sigma_{K\pi}\overline\Delta_2-\frac{(F_K/F_\pi-1)}
{\Delta_{K\pi}} \,\Biggr]\frac{\DeKP^2 }{\Sigma_{K\pi}} \,.
\end{eqnarray}
It is worth stressing that eq.~\eqn{neweqf0} is not an identity since is not
valid for arbitrary values of $F_0(0)$. As discussed in section 3 and 4,
imposing consistency between input and output values for $F_0(0)$ would
make it feasible to fix $F_0(0)$ without employing the value given in
ref.~\cite{bt03} as an input. Indeed, solving for $F_0(0)=F_+(0)$ in
eq.~\eqn{neweqf0} one explicitly finds:
\begin{eqnarray}
F_+(0)&=&\left[
1+\Delta(0)+\frac{\DeKP^2}{\Sigma_{K\pi}}
\left(\overline{\Delta}_1+\Sigma_{K\pi}\overline\Delta_2
+\frac{F_K/F_\pi-1}{\DeKP}-\frac{F_K}{F_\pi}\left[\cF_{20}'(0)+\cF_{20}''(0)
\frac{\Sigma_{K\pi}}{2}
\right]\right)\right] \nonumber \\
\label{newf0i}
&\times&\left[1+\frac{\DeKP^2}{\Sigma_{K\pi}}\left(\cF_{10}'(0)+
\cF_{10}''(0)\frac{\Sigma_{K\pi}}{2}\right)\right]^{-1} \,.
\end{eqnarray}
The values of the derivatives $\cF'_{i0}(0)$ and $\cF''_{i0}(0)$ with $i=1,2$
slightly vary over the T-matrices used in ref.~\cite{jop01}.\footnote{For the
fit 6.10K3 one has: $\cF_{10}'(0)=-3.346$, $\cF_{10}''(0)=-7.186$,
$\cF_{20}'(0)=3.335$ and $\cF_{20}''(0)=7.121$, in units of GeV$^{-2}$ and
GeV$^{-4}$ for the first and second derivatives, respectively.} Nevertheless,
this source of error is negligible compared with the uncertainties from the
rest of inputs that enter on the right hand side of eq.~\eqn{newf0i}, already
introduced in section 3. From eq.~\eqn{newf0i} one then obtains:
\begin{equation}
\label{af0f}
F_0(0) \;=\; 0.966 \pm 0.041 \,.
\end{equation}
Unfortunately, the large error of $\Delta(0)$ in eq.~\eqn{Del0}, due to lack
of a precise knowledge of the $L_i^r$ coefficients, prevents this method to be
competitive since the resulting error bar is a factor of four larger than the
one in refs.~\cite{lr84,bt03} and in eq.~\eqn{Fp0t}.

In ref.~\cite{jop01}, also different solutions for the strangeness changing
scalar  $K\pi$ and $K\eta'$ form factors were found with only one independent
solution that vanishes at infinity. They correspond to the fits 6.10K1 and 
6.11K1 of this reference. As discussed in ref.\cite{jop01}, one can pass from
the case with one independent and vanishing solution at infinity to the most
general one of eq.~\eqn{newfs}, by slightly varying three free parameters
above 1.9 GeV, giving rise to very little changes in the scattering amplitudes
for such high energies so that the same set of data is properly reproduced.
However, the one independent solution case provides us with a tighter
determination of the scalar form factors since only one unknown constant,
namely $\alpha_1$, appears.  This should result in a determination of $F_0(0)$
with smaller uncertainty than the one in \eqn{af0f}. The expressions obtained
above from \eqn{newfs} to \eqn{newf0i} for the general case can be
particularised to the one independent solution case by just equating
${\cal F}_{2i}(s)=0$, with $i=0,$ $1$. Thus, from eq.~\eqn{newf0i} one has:
\begin{eqnarray}
F_+(0)&=&\left[
1+\Delta(0)+\frac{\DeKP^2}{\Sigma_{K\pi}}
\left(\overline{\Delta}_1+\Sigma_{K\pi}\overline\Delta_2
+\frac{F_K/F_\pi-1}{\DeKP}\right)\right] \nonumber \\
\label{newf0ii}
&\times&\left[1+\frac{\DeKP^2}{\Sigma_{K\pi}}\left(\cF_{10}'(0)+
\cF_{10}''(0)\frac{\Sigma_{K\pi}}{2}\right)\right]^{-1} \,.
\end{eqnarray}

Taking into account that ${\cal F}'_{10}(0)=0.803$ GeV$^{-2}$ and 
${\cal F}''_{10}(0)=1.667$ GeV$^{-4}$ for the fit 6.10K1 and
${\cal F}'_{10}(0)=0.800$ GeV$^{-2}$ and ${\cal F}''_{10}(0)=1.661$ GeV$^{-4}$
for the fit 6.11K1, we end up with the value:
\begin{equation}
\label{af0fb}
F_0(0) \;=\; 0.979 \pm 0.009 \,.
\end{equation}
This method for fixing $F_0(0)$ is competitive with the result given in 
eq.~\eqn{Fp0t}, and even slightly more precise.  Within errors, \eqn{Fp0t},
\eqn{af0f} and \eqn{af0fb} are all found to be compatible. We can then employ
eq.~\eqn{c12c34} and eq.~\eqn{aa.6}, with
${{\cal F}}'_{20}(0)={\cal F}''_{20}(0)=0$, to fix $F_+(0)|_{C_i^r}$ with the
result
\begin{equation}
F_+(0)|_{C_i^r} \;=\; -\,0.013\pm 0.009 \,,
\end{equation}
compatible with our previous finding \eqn{Fp0Ci} and ref.~\cite{lr84}.

Let us finally note that the resulting value for $F_0(0)$ in eqs.~\eqn{Fp0t},
\eqn{af0f} and \eqn{af0fb} requires as an input the quotient $F_K/F_\pi$ that
we have fixed from ref.~\cite{lr84} to $1.22\pm 0.01$. Nevertheless, in order
to obtain this value, ref.~\cite{lr84} already employed $F_+(0)$ for the $K\pi$
system that was calculated from ${\cal O}(p^4)$ $\chi PT$, together with their
estimate of $F_+(0)|_{C_i^r}$. Therefore, our numbers for $F_+(0)$ are not
completely independent from the results of ref.~\cite{lr84}. To indicate the
dependence on $F_K/F_\pi$, using eq.~\eqn{newf0ii} we present two more results
for $F_0(0)$ for other central values of $F_K/F_\pi$, with the same uncertainty
of $\pm\,0.01$:\footnote{We can equate eqs.~\eqn{newf0i} and \eqn{newf0ii} and
then solve for $F_K/F_\pi$ in terms of $\Delta(0)$, $\overline{\Delta}_1$,
$\overline{\Delta}_2$ and the first and second derivatives at the origin of
the two kind of solutions of the scalar form factors, which have different
${\cal F}_{10}(s)$ although we have kept the same symbol. This results in
fully independent evaluations of $F_K/F_\pi$ and of $F_0(0)$ to those of
ref.~\cite{lr84}. Performing such an exercise one finds:
$F_K/F_\pi=1.19\pm 0.06 \Rightarrow F_0(0)=0.96\pm 0.05\,$. Our result for
$F_K/F_\pi$ is compatible with that of ref.~\cite{lr84} although a factor of
six less precise. One would definitely need to improve the precision in our
knowledge of $\Delta(0)$ to have more accurate results. For a hypothetical
$5\%$ error for $\Delta(0)$ the error in $F_K/F_\pi$ would turn out to be 0.2
and that for $F_0(0)$ would be 0.016.}
\begin{equation}
\frac{F_K}{F_\pi} \;=\; 1.20 \;\Rightarrow\; F_0(0) \;=\; 0.965\pm 0.009 \,,
\quad
\frac{F_K}{F_\pi} \;=\; 1.24 \;\Rightarrow\; F_0(0) \;=\; 0.993\pm 0.009 \,.
\end{equation}

\newpage

\end{document}